\documentclass[aps,pre,twocolumn,showpacs,groupedaddress]{revtex4-2}

\usepackage[latin1]{inputenc}
\usepackage[T1]{fontenc}
\usepackage{graphicx}
\usepackage{amssymb}
\usepackage{amsmath}
\usepackage{wasysym}
\usepackage{xcolor}
\usepackage{MnSymbol}
\usepackage{lettrine}
\usepackage{ulem} 

\begin{document}

\title{Penetration of a spinning sphere impacting a granular medium}

\author{D. D. Carvalho \textsuperscript{1,2}}
\author{Y. Bertho \textsuperscript{1}}
\author{E. M. Franklin \textsuperscript{2}}
\author{A. Seguin \textsuperscript{1}}

\affiliation{\textsuperscript{1} Universit\'e Paris-Saclay, CNRS, Laboratoire FAST, 91405 Orsay, France}
\affiliation{\textsuperscript{2} Faculdade de Engenharia Mec\^anica, Universidade Estadual de Campinas (UNICAMP), Rua Mendeleyev, 200 Campinas, SP, Brazil}

\begin{abstract}
We investigate experimentally the influence of rotation on the penetration depth of a spherical projectile impacting a granular medium. We show that a rotational motion significantly increases the penetration depth achieved. Moreover, we model our experimental results by modifying the frictional term of the equation describing the penetration dynamics of an object in a granular medium. In particular, we find that the frictional drag decreases linearly with the velocity ratio between rotational (spin motion) and translational (falling motion) velocities. The good agreement between our model and our experimental measurements offers perspectives for estimating the depth that spinning projectiles reach after impacting onto a granular ground, such as happens with seeds dropped from aircraft or with landing probes.
\end{abstract}

\maketitle

\section{Introduction}
\label{Sec:intro}
Granular materials, consisting of discrete, macroscopic particles, are ubiquitous in our daily lives and play a crucial role in a wide array of natural and industrial processes. From the grains of sand on a beach to the particles in pharmaceutical manufacturing, granular media are encountered across various scales and applications. Understanding their behavior, especially when subjected to external forces like impacts, is essential not only for scientific curiosity but also for solving practical engineering challenges. Impact, defined as a sudden application of force or energy, can lead to a cascade of intricate interactions among granular particles. The outcome of these interactions can vary widely, ranging from simple grain rearrangements to complex phenomena like energy dissipation, wave propagation, and even the onset of granular flows. The interest in impact within granular media spans across multiple disciplines, including physics, engineering, geology, and material science.

The penetration dynamics of a sphere in a granular medium has been extensively studied both experimentally \cite{Uehara2003, Ambroso2005, DeBruyn2004, Seguin2009}, numerically \cite{Seguin2009, Carvalho2023, Carvalho2023b}, and by modeling \cite{Goldman2008, Hinch2014, Katsuragi2007, Guo2018}. The mechanical actions involved when an object impacts a granular material are often related to the drag force. This drag force generates penetration resistance, and, for a sphere impacting a granular material, it consists of a sum of two terms. The first term is frictional, linked to the hydrostatic pressure prevailing beneath the sphere during penetration. The second term is of collisional origin, linked to the dissipation of energy by collisions between grains. However, most of those studies are limited to the case of pure translation of the sphere, with no rotational motion. Taking this rotational movement into account is important for many problems of locomotion \cite{Li2013, Zhang2014, Hosoi2015, Kumar2019, Seguin2022}, physical biology \cite{Jung2017}, and military applications \cite{Robins1745}.

Although substantially less studied in granular materials \cite{Carvalho2023, Carvalho2023b, Ye2012, Ye2015, Kumar2019}, the effect of rotation has been widely studied in Newtonian fluids. The Magnus effect is a well-known phenomenon that finds applications ranging from boat sailing to the physics of sports. In those situations, the rotation applied perpendicular to the direction of the flow creates lift forces on the object. The Magnus effect has also been observed for granular materials; however, the direction of the lifting force was found to be opposite to that normally found in viscous fluids \cite{Kumar2019}. Considering the flow of a Newtonian fluid around a sphere, the drag force, i.e., parallel to the flow direction, is not significantly affected by the rotation of the object. In particular, if the axis of rotation is aligned with the flow direction, then the drag force remains similar to the case without rotation for both high and low Reynolds numbers \cite{Gladkov2022}. Granular media exhibit a different behavior in comparison with a Newtonian fluid, their ability to modify the pressure under shear being a major feature of those materials.

In this paper, we investigate experimentally the impact dynamics of a sphere colliding with a granular bed, considering rotational effects. We observe that the rotation has an influence on the penetration dynamics, increasing the sphere penetration as it rotates faster. Additionally, we develop a first-order model that takes the effect of spinning into account to describe our experimental observations. The resulting model can be used for estimating the depth reached by spinning projectiles, with important applications in agriculture, reforestation, civil constructions, and planetary exploration.

\section{Experimental setup}
\label{Sec:setup}

\begin{figure}[t]
\centering
\includegraphics[width=6cm]{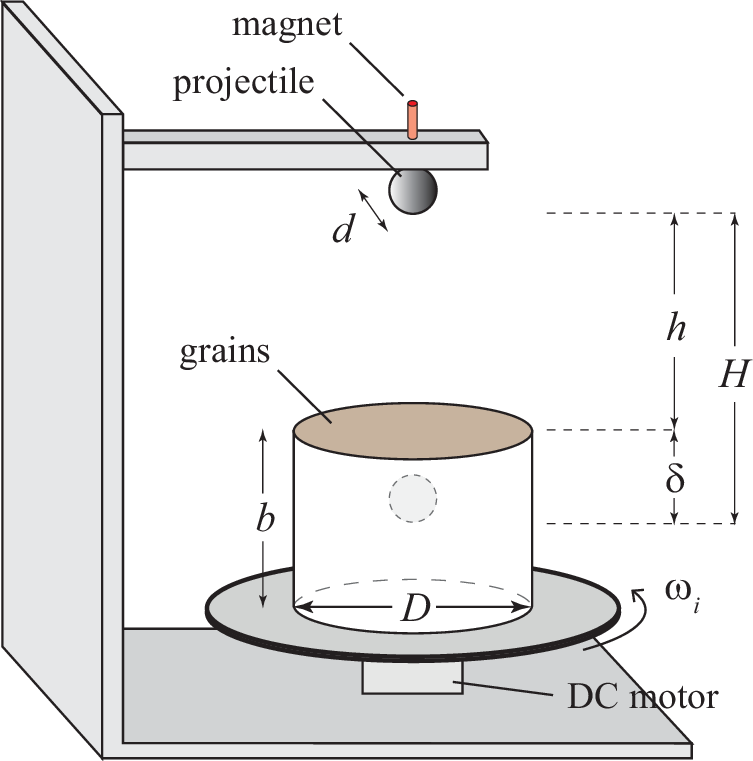}
\caption{Sketch of the experimental setup and notations introduced.}
\label{fig1}
\end{figure}
The penetration depth $\delta$ of a spherical projectile of diameter $d$ and density $\rho$ is investigated by dropping it onto a fine granular material confined in a cylindrical container of diameter $D$ and height $b$ (Fig.~\ref{fig1}). The granular medium consists of slightly polydisperse glass spheres (diameter $d_g= 1\pm 0.3$~mm and density $\rho_g\simeq 2.5\times 10^3$~kg~m$^{-3}$). To ensure the randomness of the initial conditions, the cylindrical container is overfilled with grains which are gently mixed with a thin rod before each experiment, then the free surface is flattened using a straight edge. We found that this procedure leads to reproducible measurements, with only small variations and an initial packing fraction of $\phi \approx 0.6$ (measured by weighing the container's contents).

Different materials and sizes of spherical projectiles were used to highlight the influence of the sphere density $\rho$ and diameter $d$ on the penetration depth $\delta$. For a container of diameter $D=140$~mm and height $b=90$~mm, three different metallic projectiles were used, with diameters $d=$~20, 25, and 30~mm and densities $\rho \simeq$~14920, 8160, and 7710~kg~m$^{-3}$, respectively. For a plastic projectile with diameter $d=80$~mm and density $\rho \simeq 1150$~kg~m$^{-3}$, two different cylindrical containers were used, with diameter $D=400$~mm and heights equal to $b=50$~mm and $b=180$~mm. This allows us to keep a ratio $D/d\gtrsim 5$, avoiding any confinement effects \cite{Seguin2008}. Note that for all cases studied here, the grain size $d_g$ remains much smaller than the projectile diameter $d$, with $d_g/d$ $\leq$ $0.05$.

The projectile is initially held by a magnet at a distance $h$ above the granular surface, so that it is released without any initial velocity nor rotational motion (the nonmetallic sphere is held by inserting a small magnet inside it). The impact velocity is adjusted by varying the releasing height $h$ from 2~mm to 2.17~m. Hence, the corresponding speed at impact $v_i$, given by $v_i=\sqrt{2gh}$, where $g$ is the gravitational acceleration, varies from 0.2 to 6.5~m~s$^{-1}$. Note that the projectile is released just above the center of the cylindrical container and falls along its axis.

Rather than releasing a rotating projectile to impact a nonrotating granular bed, it is more feasible to release a nonrotating projectile onto a rotating granular medium. In the absence of relatively strong centrifugal effects, these situations are equivalent, since they induce the same relative motion between the projectile and grains. Hence, the cylindrical container is placed on a rotating platform allowing its rotation around its main axis at a constant angular velocity $\omega_i$ in the range 0~--~10~rad~s$^{-1}$. In this way, the relative angular velocity between the projectile and the granular bed reaches a zero value at the end of each experiment. We note that the experiments are limited to angular velocities of the order of 10~rad~s$^{-1}$ since beyond this value, grains begin to be ejected due to the centrifugal force. In addition, at the maximum rotation rate of the system, we do not observe any surface deformation \cite{Huang2021, Irie2021}. At the end of a release experiment, the angular velocity of the container is slowly reduced until it stops completely. The penetration depth $\delta$ is then measured using a thin probe that locates the top of the projectile with an accuracy higher than 1~mm. In the following, we will denote by an index ``0'' the quantities referring to the case without any rotation ($\omega_i=0$).

\section{Impact without rotation}
\label{Sec:norotation}

\begin{figure}[t]
\centering
\includegraphics[width=\linewidth]{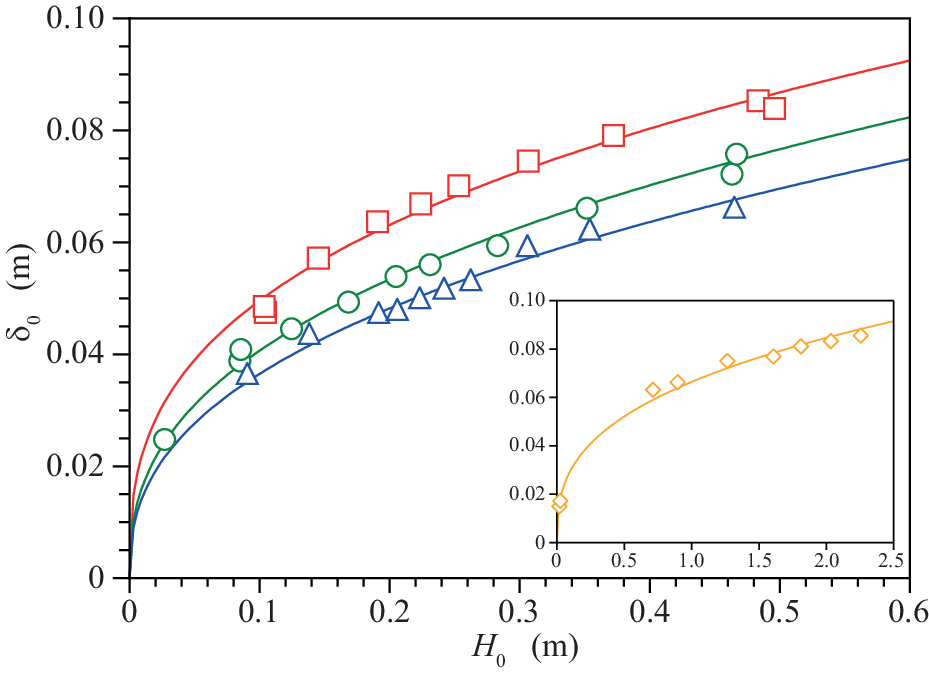}
\caption{Penetration depth $\delta_0$ as a function of the total distance $H_0=h+\delta_0$ traveled by the projectile for (\textcolor{red}{$\square$})~$d = 20$~mm, $\rho\simeq 14920$~kg~m$^{-3}$, and $55 \leq h \leq 412$~mm; (\textcolor{blue}{$\triangle$})~$d = 25$~mm, $\rho\simeq 8160$~kg~m$^{-3}$, and $54.5 \leq h \leq 400$~mm; and (\textcolor{green}{$\circ$})~$d = 30$~mm, $\rho\simeq 7710$~kg~m$^{-3}$, and $2 \leq h \leq 391$~mm. Inset: $\delta_0$ as a function of $H_0$ for (\textcolor{orange}{$\diamond$})~$d = 80$~mm, $\rho\simeq 1150$~kg~m$^{-3}$, and $3 \leq h \leq 2170$~mm. Solid lines are the best fits of experimental data following $\delta_0\propto H_0^\alpha$, with $\alpha=$ 0.35, 0.40, 0.39, and 0.35 for increasing projectile sizes.}
\label{fig2}
\end{figure}

Figure~\ref{fig2} displays the penetration depth $\delta_0$ as a function of the total distance $H_0=h+\delta_0$ traveled by the projectile, for a motionless reservoir ($\omega_i=0$) and four different projectiles of different diameters $d$ and densities $\rho$.
As expected, and as already observed in several studies, the bigger and/or denser the projectile, the deeper it sinks into the granular medium \cite{Seguin2008}. Moreover, the greater the drop height, the deeper the penetration depth, with a power-law dependence close to $\delta_0\propto H_0^{0.4}$ in good agreement with the empirical power laws proposed in previous studies \cite{Uehara2003, Ambroso2005, DeBruyn2004, Seguin2008}.

\begin{figure}[t]
\centering
\includegraphics[width=\linewidth]{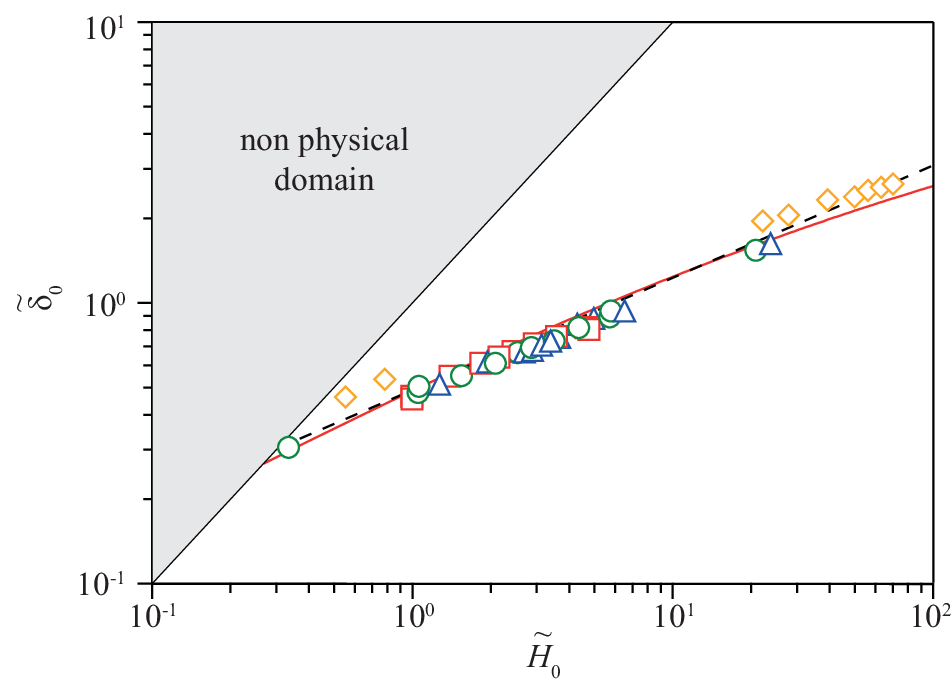}
\caption{Nondimensional penetration depth $\tilde{\delta}_0$ as a function of the nondimensional total distance $\tilde{H}_0=\tilde{h}+\tilde{\delta}_0$ traveled by the projectile in $\log$-$\log$ scales for the same data as in Fig.~\ref{fig2}. (-~-~-)~Best fit of the data, with $\tilde{\delta}_0= 0.49~\tilde{H}_0^{0.4}$. (\textcolor{red}{---})~Solution of Eq.~(\ref{eq3}), with $K_{v}=0.39$ and $K_z=7.2$.}
\label{fig3}
\end{figure}

Let us then write the equation describing the dynamics of the projectile. The projectile experiences its own weight and a force resulting from its interactions with the granular medium, the latter having collisional and frictional origins that are proportional to $v^2$ and $z$, respectively \cite{Katsuragi2007}. Hence, the equation of motion for the projectile can be written as
\begin{equation}
\rho\frac{\pi d^3}{6}\frac{d^2z}{dt^2}=\frac{\pi d^3}{6}\rho g - K_v\rho_g\phi d^2v^2 - K_z \rho_g\phi gd^2z,
\label{eq1}
\end{equation}
where $K_v$ and $K_z$ are coefficients. Based on previous theoretical studies \cite{Guo2018}, we can define a dimensionless depth $\tilde{z}$ and a dimensionless time $\tilde{t}$ as
\begin{equation}
\tilde{z}=\frac{6\rho_g\phi}{\pi \rho d}z \qquad \mathrm{and}\qquad \tilde{t}=\left( \frac{6\rho_g\phi g}{\pi \rho d}\right)^{1/2}t.
\label{eq2}
\end{equation}
This allows one to build the dimensionless penetration depth $\tilde{\delta}_0$ and the dimensionless total distance $\tilde{H}_0=\tilde{h}+\tilde{\delta}_0$, which are plotted in Fig.~\ref{fig3}. We observe that all the experimental data collapse remarkably well on a single curve following the scaling law $\tilde{\delta}_0 =A{\tilde{H}_0}^\alpha$, with $A\simeq 0.49$ and $\alpha\simeq 0.4$ (dashed line in Fig.~\ref{fig3}).

By using the same scaling from Eqs.~(\ref{eq2}), Eq.~(\ref{eq1}) can be made dimensionless,
\begin{equation}
\frac{d^2 \tilde{z}}{d \tilde{t}^2}= 1 - K_v \left(\frac{d \tilde{z}}{d \tilde{t}}\right)^2 - K_z \tilde{z},
\label{eq3}
\end{equation}
and be solved numerically with the initial conditions $z=0$ and $dz/dt=v_i=\sqrt{2gh}$ at $t=0$, for the position and the impact velocity of the projectile, respectively, i.e., in dimensionless notations $\tilde{z}=0$ and $\tilde{v}_i=\left(6\rho_g\phi/\pi\rho g d\right)^{1/2}v_i$. The final penetration depth $\tilde{\delta}_0$ corresponds to the depth where the velocity vanishes, the characteristic time of the penetration linked to the impact being typically $T_v=\left(6\rho_g \phi g/\pi \rho d\right)^{-1/2}$. Integrating this equation with the values $K_v=0.39$ and $K_z=7.2$ leads to the red solid curve shown in Fig.~\ref{fig3}, which is very close to the empirical power-law fit (dashed line in the same figure). A large range of variations is reported in the literature for the $K_v$ and $K_z$ coefficients, depending on material properties (density, friction coefficient, shape) and/or packing fraction $\phi$ \cite{Kang2018, Katsuragi2013, Guo2018}. In particular, note that we expect these coefficients to increase with $\phi$, leading to a decrease of the penetration depth $\tilde{\delta}_0$ with $\phi$ \cite{Umbanhowar2010, Carvalho2023}. Since our experiments are performed with the same grains and at approximately the same packing fraction, we decided to use a single pair of coefficients $K_v$ and $K_z$. We point out that the values obtained for both coefficients are in the same order of magnitude as compared to other similar studies \cite{Katsuragi2007, Pacheco2011, Katsuragi2013, Hinch2014, Guo2018}.

\section{Impact with rotation}
\label{Sec:rotation}

\subsection{Experimental and numerical results}
In the reference frame of the tank, the spherical projectile has an initial spin $\omega_i$. This rotational velocity can be compared to the impact velocity using the velocity ratio $\nu$:
\begin{equation}
\nu=\frac{\omega_i\, d}{v_i}.
\end{equation}
Note that $\nu$ can be interpreted as the square root of the ratio between angular and translational kinetic energies.

\begin{figure}[t]
\centering
\includegraphics[width=\linewidth]{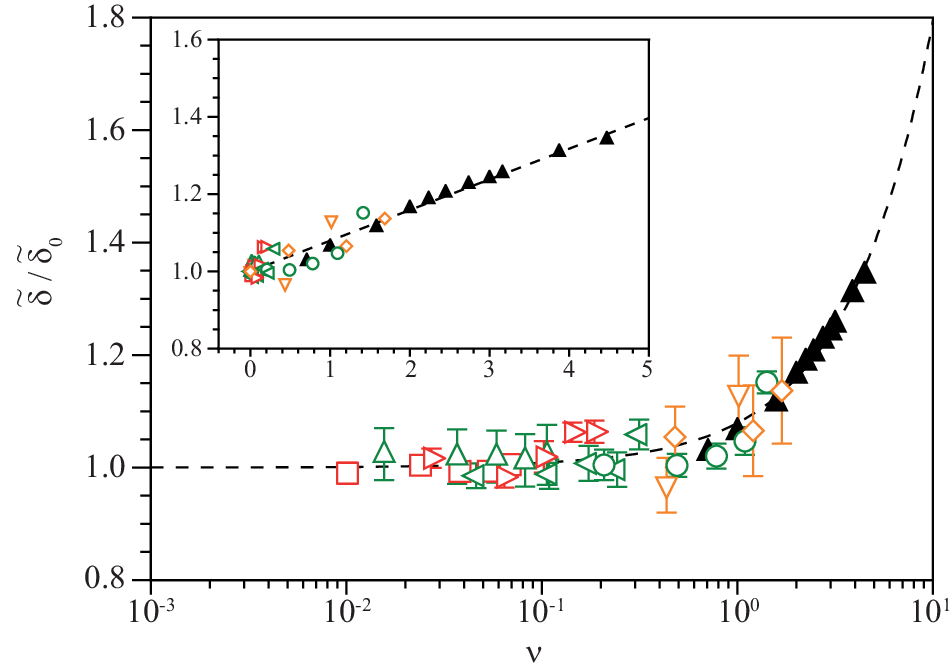}
\caption{Nondimensional penetration depth $\tilde{\delta}$ divided by the nondimensional penetration depth without rotation $\tilde{\delta}_0$, as a function of the nondimensional velocity ratio $\nu = \omega_i\, d/v_i$, for the following configurations: $d=20$~mm and (\textcolor{red}{$\triangleright$})~$h=57$~mm, (\textcolor{red}{$\square$})~$h=412$~mm ; $d=30$~mm and (\textcolor{green}{$\circ$})~$h=2$~mm, (\textcolor{green}{$\triangleleft$})~$h=45$~mm, (\textcolor{green}{$\triangle$})~$h=391$~mm ; $d=80$~mm and (\textcolor{orange}{$\diamond$})~$h=3$~mm, (\textcolor{orange}{$\triangledown$})~$h=8$~mm. ($\blacktriangle$)~Numerical simulations from Ref.~\cite{Carvalho2023}. (-~-~-)~Best fit of both experimental and numerical data with $\tilde{\delta}/\tilde{\delta}_0=1+0.08\,\nu$. Inset: Same data in linear plot.}
\label{fig4}
\end{figure}
Figure~\ref{fig4} displays the evolution of the relative penetration depth $\tilde{\delta}/\tilde{\delta}_0$ as a function of the velocity ratio $\nu$ (note that $\tilde{\delta}/\tilde{\delta}_0$ is also simply $\delta/\delta_0$). We observe that the data collapse on a single curve, as well as a slight increase in penetration depth with rotation. Indeed, within the experimental range of $\nu$, the relative rotation between the projectile and the grains leads to variations in the penetration depth of up to approximately 15\%.

To enhance the experimental data, we supplement them with numerical results extracted from a previous study obtained from three-dimensional discrete element method computations \cite{Carvalho2023}. Their granular medium consisted of spheres ($d_g=1\pm 0.4$~mm, $\rho_g=2600$~kg~m$^{-3}$) confined in a cylindrical container ($D=125$~mm, $b=76.5$~mm), leading to a granular packing fraction of $\phi\simeq 0.554$. The projectile ($d=15$~mm, $\rho=7865$~kg~m$^{-3}$) was released from a height $h=10$~mm, with an angular velocity $\omega_i$ that varied from 0 to 418~rad~s$^{-1}$. The results of these simulations are plotted in Fig.~\ref{fig4} (dark symbols) and complement our experimental findings at large $\nu$, in a range of parameters that are difficult to access experimentally, since if $\omega_i$ is too large, then the grains tend to be ejected due to centrifugal forces. A very good agreement between experimental and numerical data is observed, with an overlap zone around $\nu\simeq 1$. For rotating projectiles impacting a granular bed \cite{Carvalho2023}, part of the rotational kinetic energy of the projectile agitates the bed, helping to dislodge more grains and excavate it, leading to an increase in the penetration depth.

The inset in Fig.~\ref{fig4} shows the same data in linear plot and suggests that the relative penetration depth $\tilde{\delta}/\tilde{\delta_0}$ increases linearly with the velocity ratio $\nu$, following a law of the form $\tilde{\delta}/\tilde{\delta}_0 =1+B\nu$, with $B\simeq 0.08$. Finally, considering that $\tilde{\delta}_0 = A \tilde{H}_0^\alpha$, the data are well described by the general fit,
\begin{equation}
\tilde{\delta}=A \tilde{H}_0^\alpha \left(1+ B\nu \right),
\label{eq5}
\end{equation}
where $A\simeq 0.49$ and $\alpha\simeq 0.4$ come from the case without rotation ($\nu \rightarrow 0$). The penetration depth $\delta$ of the projectile, expressed in terms of impact velocity $v_i$ and rotation velocity $\omega_i$, writes
\begin{equation}
\delta=Ad\left(\frac{\pi\rho }{6\rho_g\phi}\right)^{1-\alpha}\left (\frac{v_i^2}{2gd}+\frac{\delta_0}{d}\right )^\alpha \left (1+B\frac{\omega_i\, d}{v_i}\right ).
\label{eq6}
\end{equation}
We note that the scaling drawn in Fig.~\ref{fig4} is not a master curve since Eq.~(\ref{eq6}) cannot be written in a separable form with $v_i$ and $\omega_i$. Consequently, $\tilde{\delta}/\tilde{\delta}_0$ is still a function of both variables $v_i$ and $\omega_i$. However, the dependence of $\tilde{\delta}/\tilde{\delta}_0$ with $v_i$ is quite soft, which is why we observe that all the data collapse on a single curve for the range of $\omega_i$ and $v_i$ explored.
\begin{figure}[t]
\centering
\includegraphics[width=\linewidth]{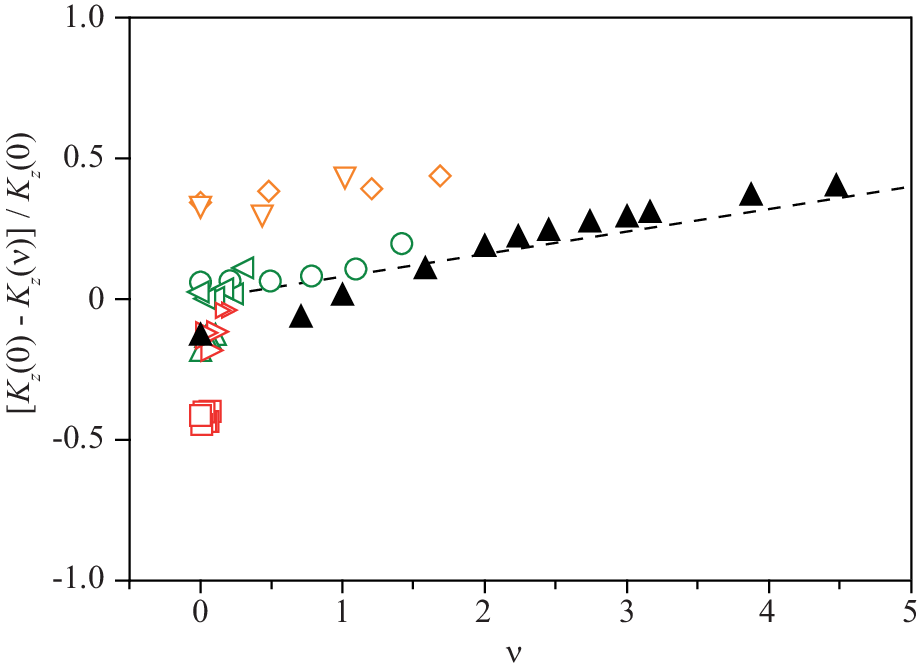}
\caption{$[K_z(0)-K_z(\nu)]/K_z(0)$ as a function of the nondimensional ratio velocity $\nu = \omega_i \,d/v_i$. Symbols correspond to the same data as in Fig.~\ref{fig4}. Numerical data have been integrated with $K_z(0)=26.84$ and $K_v=0.68$ corresponding to values for the nonrotating projectile in numerical simulations. (-~-~-)~ Fit of both experimental and numerical data with $[K_z(0)-K_z(\nu)]/K_z(0) = 0.08\,\nu$.}
\label{fig5}
\end{figure}

\subsection{Model update and discussion}
We observed in Fig.~\ref{fig4} that the penetration depth varies continuously across the different angular velocities (spins), for all initial heights and angular velocities tested experimentally and numerically \cite{Carvalho2023}. This suggests that the dynamics of the projectile should be continuous across different initial spins as it penetrates the granular bed. Therefore, we aim at modifying Eq.~(\ref{eq3}) by incorporating the contribution of the projectile spin.

The increase in penetration depth with the initial spin indicates that rotation enhances the fluidization of part of the granular medium around the projectile. As a result, the pressure beneath the projectile decreases, with its angular velocity decreasing over a characteristic time $T_\omega$ due to a resisting torque exerted by the granular medium on the sphere. This resisting torque corresponds to a local solid friction force per unit area between the grains and the sphere, $\tau_f$, independent of the spin and geometrical properties of the projectile. So a reasonable scale for this resistive torque should be $d^3 \tau_f$. Carrying out a dimensional analysis for the conservation of angular momentum leads to $\rho d^5 \omega_i/T_\omega \sim d^3 \tau_f$, i.e., $T_{\omega} \propto \rho d^2 \omega_i $. As rotation and vertical penetration are decoupled in our experiments, there is no obvious reason for the two corresponding timescales $T_\omega$ and $T_v$ to be of the same order of magnitude. For example, for low rotational velocity, the characteristic time $T_\omega$ is such as $T_\omega \ll T_v$, so that the rotation of the projectile can be neglected. Conversely, for high rotational velocity ($T_{\omega} \gg T_v$), the projectile can keep on rotating while final penetration is reached. It is then reasonable to assume that the resisting torque arises from contact between the grains and the projectile and is therefore frictional rather than collisional in origin. Hence, we will suppose that only the coefficient $K_z$ will be affected by the rotation of the projectile $\omega(t)$, while the coefficient $K_v$ will not depend on it. Let us consider two limiting cases: (\textit{i})~The first one is the regime where the impact velocity $v_i$ dominates and leads to a penetration time $T_v$ much greater than the characteristic time for spinning $T_\omega$. In this situation, rotation will have no effect on penetration depth. (\textit{ii})~The second one is the regime where the characteristic time due to rotation $T_\omega$ is much greater than that due to translation $T_v$. In this case, the angular velocity can be considered constant over time, so that $\omega(t)\sim \omega_i$. Both extreme cases are well describe by the relevant dimensionless number $\nu=\omega_i\, d/v_i$. Consequently, in first approximation, the effect of rotation can be modeled by defining a function $K_z(\nu)$. Let us set
\begin{equation}
K_z(\nu)=K_z(0) [1-\lvert \chi(\nu) \rvert].
\end{equation}
Note that the absolute value allows to overcome the direction of rotation of the projectile.

Let us determine $\chi(\nu)$ in our experiments. To do so, we set $K_z(0)=7.2$ and $K_v=0.39$ (same values as for the nonrotating projectile) and determine $\chi$ for solving Eq.~(\ref{eq3}) continuously across different values of $\nu$. Figure~\ref{fig5} shows the evolution of $[K_z(0)-K_z(\nu)]/K_z(0)$ as a function of $\nu$ for both experimental and numerical data. Although with a significant dispersion of data at low values of $\nu$, we observe that the coefficient $K_z$ decreases with $\nu$, which is consistent with the fact that the penetration depth $\delta$ increases with the angular velocity. This decrease is rather linear, so that we can define the function $\chi$ such that $\chi(\nu)=C\nu$ with $C\simeq 0.08$ to fit the data (dashed line in Fig.~\ref{fig5}). In this way, we are now able to modify the initial model [Eq.~(\ref{eq3})] for taking rotation effects into account, leading to:
\begin{equation}
\frac{d^2 \tilde{z}}{d \tilde{t}^2}= 1 - K_v \left(\frac{d \tilde{z}}{d \tilde{t}}\right)^2 - K_z(0) (1-\lvert C\nu \rvert) \tilde{z}.
\label{eq8}
\end{equation}
We observe a discrepancy between the experimental data for the highest diameter ($d=80$~mm) and the dashed line on Fig.~\ref{fig5}. These experiments always exhibit a penetration depth smaller than the projectile diameter ($\delta /d < 1$). Since the model is based on the assumption that the projectile is a point object, it is not precise enough to capture such limiting cases. It should be noted, however, that the trend remains consistent, since these data follow a curve parallel to the dashed line. In addition, we believe the dispersion observed for the densest sphere ($d=20$~mm) to be associated with the fact that $K_z(0)$ and $K_v$ are supposed to be constant (although the data are well described by one single pair of coefficients, as seen in Fig.~\ref{fig3}). Moreover, the linear model proposed in Eq.~(\ref{eq8}) works only for cases where $|\chi({\nu})| < 1$, ensuring energy dissipation.

\begin{figure}[t]
\centering
\includegraphics[width=\linewidth]{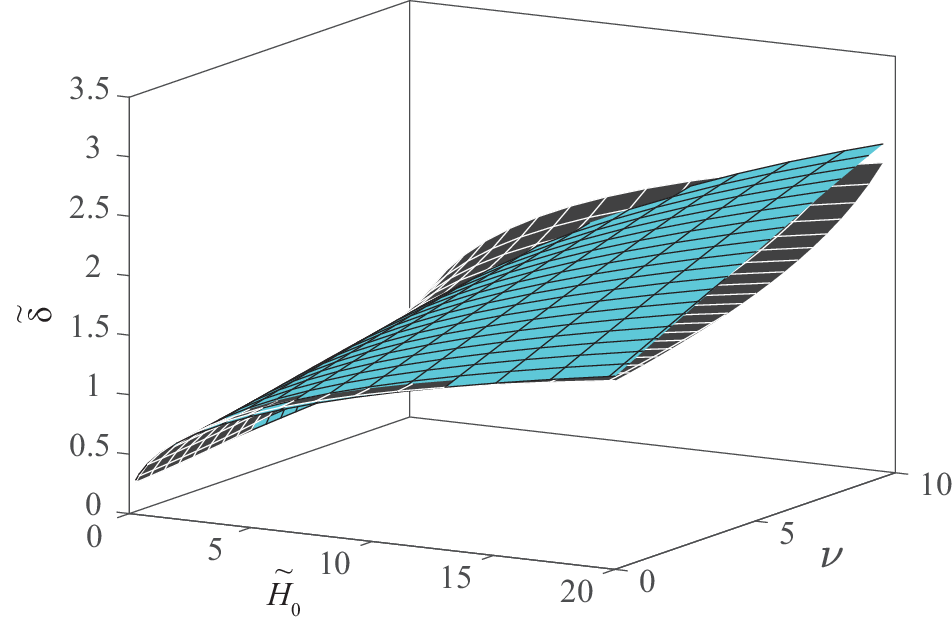}
\caption{Nondimensional penetration depth $\tilde{\delta}$ as a function of the dimensionless total distance $\tilde{H}_0$ and the nondimensional velocity ratio $\nu$. Black surface corresponds to the solution of Eq.~(\ref{eq8}). Blue surface corresponds to the scaling law (\ref{eq5}).}
\label{fig6}
\end{figure}

The resolution of Eq.~(\ref{eq8}) generates the black surface displayed in the three-dimensional graphic of Fig.~\ref{fig6}, which describes the penetration depth $\tilde{\delta}(\tilde{H}_0,\nu)$ as a function of the total height and the velocity ratio. Figure~\ref{fig6} also shows a blue surface that represents the scaling law described by relation~(\ref{eq5}). We can see that both surfaces are very close to each other, showing that the proposed model correctly captures the observed scaling law.

\section{Conclusion}
\label{Sec:Conclusion}
In this study, we have shown that rotational velocity increases the penetration depth of a projectile impacting a granular material. However, the effect of rotation on the penetration depth is of a lower order than that of impact velocity. It is possible to take the effect of rotation $\nu$ into account in the penetration depth $\tilde{\delta}$ by modifying the usual scaling law for penetration depth without rotation $\tilde{\delta}_0$. The new scaling law is an affine relation that reads $(\tilde{\delta}-\tilde{\delta}_0)/\tilde{\delta}_0 \sim \nu$. Moreover, by implementing rotation effects, we have adapted the dynamic equation for penetration of a projectile into granular materials. The influence of rotation $\nu$ essentially reduces the frictional drag term (proportional to depth). The resolution of this updated equation leads to a solution $\tilde{\delta}(\tilde{H}_0, \nu)$ represented by a nonplanar surface very close to the aforementioned new scaling law.

Despite the progress achieved, some important questions remain to be further investigated experimentally, such as how crater morphology is affected by projectile rotation as well as the case of oblique impact of rotating projectiles. Furthermore, it would be interesting to propose a more complex model that fully takes the deceleration of the projectile spin $\omega(t)$ into account during the collisional process. Nevertheless, the model proposed in this article can be used to estimate the depth reached by spinning projectiles, with important industrial, geotechnical and environmental applications. Overall, our results represent a new step for understanding the mechanics of impact in granular media.

\begin{acknowledgments}
The authors thank B. Darbois~Texier and M. Rabaud for fruitful discussions. We are grateful to J.~Amarni, A.~Aubertin, L.~Auffray, C.~Manquest, and R.~Pidoux for the development of the experimental setup. The authors are grateful to the S\~ao Paulo Research Foundation FAPESP (Grants No. 2018/14981-7, No. 2020/04151-7, and No. 2022/12511-9) for financial support.
\end{acknowledgments}

\bibliography{biblioSpinImpact}

\end{document}